\documentclass[doublespacing]{narfront}
\usepackage{graphicx,xspace,units,subfigure}
\usepackage{float}
\usepackage{amsmath}
\usepackage{amssymb}
\usepackage[normalem]{ulem}
\usepackage{color,soul} 
\sethlcolor{red} 

\newcommand{\seqc}{$\mbox{seq}_{\mbox{c}}$}
\newcommand{\seqone}{$\mbox{seq}_1$}
\newcommand{\seqtwo}{$\mbox{seq}_2$}
\newcommand{\seqthree}{$\mbox{seq}_3$}
\begin{document}

\begin{frontmatter}
\title{Predicting Transcription Factor Specificity with All-Atom Models}
\author[label1]{Sahand Jamal Rahi}
%\ead{sjrahi@mit.edu}
\author[label2]{Peter Virnau\corauthref{cor}}
\corauth[cor]{Corresponding author.}
%\ead{kardar@mit.edu}
\ead{virnau@uni-mainz.de}
\author[label1,label3]{Leonid A. Mirny}
%\ead{leonid@mit.edu}
\and
\author[label1]{Mehran Kardar}
%\corauth[cor]{Corresponding author.}
%\ead{kardar@mit.edu}

\address[label1]{Massachusetts Institute of Technology, Department of Physics, 77 Massachusetts Avenue, Cambridge, MA 02139, USA}
\address[label2]{Staudinger Weg 7, Institut f\"ur Physik, 55099 Mainz, Germany}
\address[label3]{Harvard-MIT Division of Health Sciences and Technology, Massachusetts Institute of Technology, 77 Massachusetts Ave., Cambridge, MA 02139, USA}
%\email{kardar@mit.edu}

\begin{abstract} 
\textbf{ The binding of a transcription factor (TF) to a DNA operator
site can initiate or repress the expression of a gene. Computational
prediction of sites recognized by a TF has traditionally relied upon
knowledge of several cognate sites, rather than an \emph{ab initio}
approach.  Here, we examine the possibility of using structure-based
energy calculations that require no knowledge of bound sites but
rather start with the structure of a protein-DNA complex.  We study
the PurR \emph{E. coli} TF, and explore to which extent atomistic
models of protein-DNA complexes can be used to distinguish between
cognate and non-cognate DNA sites. Particular emphasis is placed on
systematic evaluation of this approach by comparing its performance
with bioinformatic methods, by testing it against random decoys and
sites of homologous TFs. We also examine a set of experimental
mutations in both DNA and the protein.  Using our explicit estimates
of energy, we show that the specificity for PurR is dominated by
direct protein-DNA interactions, and weakly influenced by bending of
DNA.}

%We study the binding of transcription factor PurR to DNA. We compare
%ab initio specificity predictions based on all-atom models with
%bioinformatics predictions based on sequence similarity. We show that
%the specificity is predominantly due to protein-DNA interactions
%allowing us to predict the consensus sequence easily. Using binding
%energies we go on to score the binding of close mutants of PurR to DNA
%sequences, which is out of reach for bioinformatics tools. The results
%are compared to experimental data.
\end{abstract} 


\end{frontmatter}

\bibliographystyle{nar}

%\maketitle

%%%%%%%%%%%%%%%%%%%%%%%%%%%%%%%%%%%%%%%%%%%%%%%%%%%%%%%%%%%%%%%%%%%
\section{Introduction}\label{sec:Introduction}

Binding of cognate sites of DNA is central to many essential
biological processes. Most of DNA-binding proteins have the ability to
recognize and tightly bind cognate DNA sequences (sites). To find
sites bound by a particular DNA-binding protein, one needs to
calculate the free energy of binding for the protein and possible DNA
sites and then select sites that provide sufficiently low binding
energy. A widely used approach to find sites for a DNA-binding protein
is to assume a form of the energy function, infer its parameters, and
to then calculate the energy for all sites in a genome.  To infer
parameters one needs to have a set of known sites bound by the
protein. Given these known sites, the parameters are inferred using
either a widely used Berg-von Hippel approximation~\cite{BvH1986}, or
by other recently proposed methods~\cite{Sengupta,Kinney,Foat06}{.}
This constitutes a physical basis for many widely used bioinformatics
techniques that rely on a particular form of the energy function known
as a position-specific weight matrix (PWM).

All these methods require {\it a priori} knowledge of the sites (or at
least longer sequences containing these sites) bound by the
protein. This data is available for only a small number of DNA-binding
proteins. For many DNA-binding proteins, however, their sequence
of amino-acids is well known. Sufficiently high evolutionary
conservation of DNA-binding domains, and the availability of crystal
structures for many of them, makes it possible to construct 3D models
for a broad range of DNA-binding proteins. Can such protein structures be
used to predict sites recognized by a DNA-binding protein?
The basic procedure for structure-based methods
is to compute the binding energy of the protein-DNA complex.  
The structure of the complex for {\em an arbitrary DNA sequence} can be modeled by
replacing (``mutating'') the DNA sequence in the protein structure
containing its cognate site, followed by energy minimization and/or molecular
dynamics to allow the protein-DNA complex to adjust to the new DNA
sequence. After several minimization steps the interaction energy can
be calculated using either standard molecular mechanics force fields
like AMBER~\cite{parm99} or CHARMM~\cite{charmm27} with an implicit
solvent,
%\cite{Paillard04-1,Paillard04-2,Flanagan05,Endres04,Endres06}.
or a knowledge-based force field optimized for the particular complex~\cite{Donald07}.
%\cite{Morozov05,Thayer02,Gromiha04,Havranek04,Kono99,Mandel-Gutfreund01,Mandel-Gutfreund98,Selvaraj02}

Several recent studies have significantly elaborated upon the above procedure.
Lafontaine and Lavery, for example, pioneered a very efficient process
termed ADAPT~\cite{Lafontaine01} in which they replace the DNA in the
structure by a `multicopy' or `average' piece of DNA.  The structure
is only minimized once after which the energy of the complex is
measured for all possible DNA sequences in place of the average
piece. From this, only the energy of the unbound DNA must be
subtracted. The unbound protein energy is the same for all DNA
sequences and hence irrelevant for comparisons.  This approach is so
efficient that all possible sequences ($4^{N}$ for $N$ bases) for short DNA operator
sites can be evaluated.  Their results successfully identify the
experimental consensus sequence for a variety of DNA-binding 
proteins~\cite{Paillard04-1,Paillard04-2}, and the ordering of binding free
energies for DNA point mutations in several complexes~\cite{Paillard04-1}.  
In this context, it was also noted that the actual binding energy computed via 
minimizations is incorrect and cannot be compared to experiments quantitatively.

Endres, Schulthess, and Wingreen allowed protein side chains to
explore rotamer conformations in their study of
Zif268~\cite{Endres04}.  Interestingly, the agreement with experiments
becomes worse when rotamers are considered, which points to a
potential bias of the approach towards sequences similar to the one on
which the underlying experimental structure is based. Morozov et
al.~\cite{Morozov05} predict binding affinities using energy
measurements as well, they keep their structures rigid or allow them
to relax and compare the two approaches. However, instead of
considering their binding energies to be approximately equal to free
energies as we do, they fit their energies to a few experimentally
known free energies. They assign different weights to the energies
involved, e.g., the Lennard-Jones or the electrostatic energy, and
optimize the weights so that the sum matches the free energy. They
proceed to study several transcription factors and even find consensus
sequence logos for two transcription factors whose structures they
construct by homology modeling. In recent work, Donald et
al.~\cite{Donald07} focus on direct protein-DNA interactions. They
study and compare a number of potentials and propose some that
outperform the standard Amber potential.  All these efforts represent
pioneering work in the emerging field of structure-based predictions
of transcription factor specificity.

Here we explore whether widely available molecular dynamics (MD) force
fields can be used to calculate the binding free energy from all-atom
models of the protein-DNA complex. In contrast to some of the previous
studies, we (i) assess the power and limitations of the method in
dealing with the roughly $10^6$ decoy sites of bacterial genomes
(by computing binding energies for representative mutations and
assembling an energy-based weight matrix (EBWM), which is then used
for the task);
(ii) explore whether energy-minimization methods utilizing MD
force fields can predict protein-DNA binding when DNA sites, or the
protein, are mutated.

For our study we focus on the purine repressor, PurR, from {\it
E.~coli}, a well-characterized transcription factor with more than 20
known sites in the genome.  The purine repressor is a member of the
sizable LacI family, which is often regarded as a model system for
transcription regulation. The abundance of both
experimental~\cite{PurR_experiments} and
bioinformatics~\cite{Mironov99} data make this an ideal target for
testing structure-based prediction techniques, and to study their
assets and drawbacks.

We demonstrate that generic molecular dynamics tools predict favorable
binding energies for known cognate sites.  To quantify the power and
limitations of this approach we investigated the following: (i) Can we
recognize the cognate sites from a large set of decoys, and estimate
the number of false positives?  (ii) How does the performance in the
above test compare with that of a {\em motif} obtained from the set of
cognate sites by bioinformatic methods?  By calculating binding
energies we can also answer the following questions which are not
addressable by bioinformatic means: (i) What is the relative
importance to recognition of direct binding energies to indirect
factors such as DNA bending?  (ii) Can the computed results for
$\Delta G_\text{binding} $ of mutations in DNA, and more importantly
in the protein, be compared to experiment?\footnote{{Bioinformatics
data can also be converted to compute} $\Delta G_\text{binding} $
{for DNA, but not protein, mutations as in Refs.}
\cite{BvH1986,Sengupta,Kinney,Foat06}.}

To test the ability of the force field to discriminate
between cognate sites and random decoys, we developed a procedure
to speed up calculations and the screening of many sites. 
We find that a single cognate site can be
discriminated from about 7000 random decoy sites. While such
performance is impressive, it is insufficient to detect sites from the
whole bacterial genome. 
In the comparisons of our results with
experimental binding free energies for DNA and amino acid point
mutations, we obtain the correct order of binding free
energies of the mutants.

%%%%%%%%%%%%%%%%%%%%%%%%%%%%%%%%%%%%%%%%%%%%%%%%%%%%%%%%%%%%%%%%%%%
\section{Materials and Methods}\label{sec:Methods}
%\subsection{Procedure}

%% \begin{figure}[h]
%% \vspace{0.5cm}
%% \centering
%% \includegraphics[width=7cm]{ProtDNA.eps}
%% \caption{PurR protein headpiece bound to its consensus sequence
%% DNA. This structure \cite{Glasfeld99} serves as the basis of our
%% study. The DNA base pairs or the protein amino acids in this structure
%% are mutated on the computer and the effects on the binding energy
%% measured. Blue and red: protein chains, orange and grey: DNA.}
%% \label{fig:ProtDNA}
%% \end{figure}

The change in free energy due to protein-DNA interactions can be
decomposed as 
\begin{equation}
\begin{split}
G_{\text{binding}} & = G_{\text{protein-DNA complex}} \\
                   & - G_{\text{free DNA}} \\
                   & - G_{\text{free protein}}
\label{eq:Gbinding}
\end{split} ~.
\end{equation}
Clearly, $G_{\text{binding}}$ depends on both the particular DNA
sequence and the protein. In order to simplify the problem from a
computational point of view, it is often assumed that the differences
in $G_{\text{binding}}$ for two different DNA sequences are dominated
by differences in enthalpy. Entropic contributions are usually ignored
since the entropy losses upon binding for both the fragment of DNA and
the protein are likely not to depend significantly on the DNA sequence; hence
\begin{equation}
\label{eq:Ebinding}
\begin{split}
G_{\text{binding}} & \approx E_{\text{binding}} \\
                   & = E_{\text{protein-DNA complex}}\\
                   & - E_{\text{free (straight) DNA}}\\
                   & - E_{\text{free (unbound) protein}}
\end{split} ~.
\end{equation}
Furthermore, if DNA sequences bound by the same protein are compared
and $\Delta E(\text{DNA1},\text{DNA2})_{\text{binding}} =
E(\text{DNA1},\text{Protein})- E(\text{DNA2},\text{Protein})$ is of
interest, the term $E_{\text{free (unbound) protein}}$ cancels out.

\begin{figure}[h]
\vspace{0.5cm}
\centering
\includegraphics[width=0.75\columnwidth]{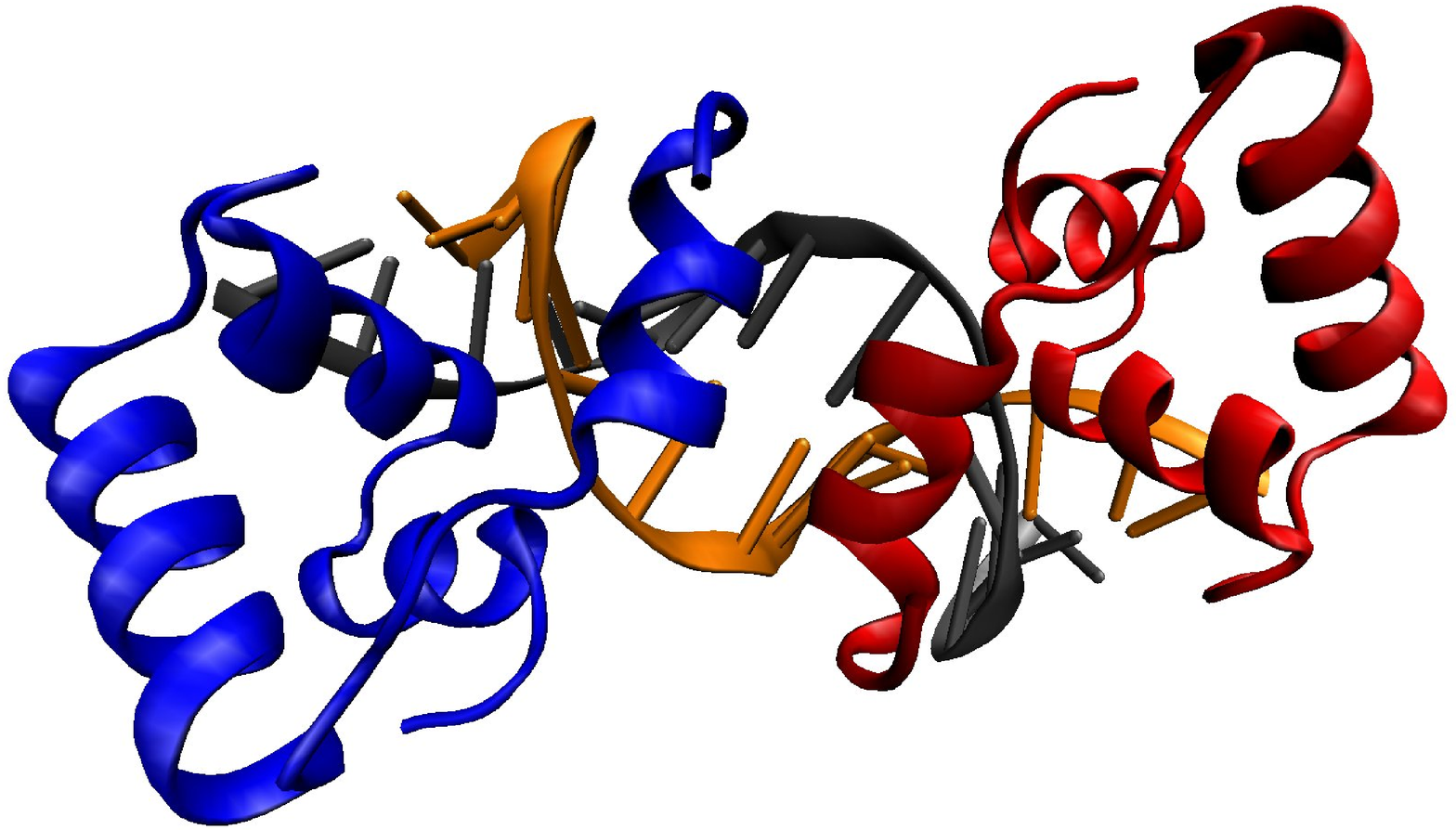}
\caption{PurR protein headpiece bound to its consensus sequence
DNA. This structure \cite{Glasfeld99} serves as the basis of our
study. The DNA base pairs or the protein amino acids in this structure
are mutated on the computer and the effects on the binding energy
measured. Blue and red: protein chains, orange and grey: DNA.}
\label{fig:ProtDNA}
\end{figure}

The energies of the molecules were measured after minimizing the
energy of their structures using the AMBER software package, its force
field, and an implicit water model.
The reference structure in this study is 1qpz~\cite{Glasfeld99}, a wild-type 
PurR structure bound to DNA.
The sequence of the DNA is also the consensus sequence obtained in the
bioinformatics study of Ref. \cite{Mironov99} and we shall thus refer
to it as the consensus sequence.
The structure, depicted in Fig. \ref{fig:ProtDNA}, was reduced to its
60 amino acid headpiece, and the DNA was trimmed to the 16 base pair
consensus sequence.  The first amino acid is missing and was not
inserted artificially. The reference for straight DNA was taken and
trimmed from the first model of the non-cognate LacI-DNA binding
complex 1osl~\cite{Kalodimos04}.
DNA sequences were exchanged with the 3DNA computer
application~\cite{3dna}. The experimental {DNA} backbone remained
in place, only the base pairs were replaced. The free DNA molecule
obtained in this manner deviates from a `perfect' B-DNA molecule by
about $1~\unit{\AA}$ RMS.
{While the experimentally derived straight DNA molecule was
  preferred over one with average coordinates, this choice had no
  significant impact on our results. The standard deviation of the
  energy difference between the canonical B-DNA structures and ours is
  merely $1.2~\unit{kcal/mol}$ for the 50 random DNA sequences that
  were used (see below). Also, for example, the linear correlation
  coefficient of $-0.6$ between bioinformatics scores and binding
  energies for the random sequences discussed below does not change at
  all.}
Protein mutants were generated with the Mutator 1.0 plugin built into
VMD~\cite{vmd}. The software uses psfgen to build a new side chain
from pre-defined parameters for the CHARMM force field; this structure
is not relaxed further by VMD. But the mutated side chain assumes a
low-energy conformation during energy minimization because -unlike the
original residues- mutated residues were not constrained to remain
close to the coordinates of the original structure (see below).
For the study of DNA point mutations the respective
structures 1qp0, 1qp4, 1bdh, 1qqb, 1qp7, 1qqa from
Ref.~\cite{Glasfeld99} were used in addition to 1qmz.  We applied
psfgen to combine and prepare structures for minimization.

It should be noted that the conformational energy of the free unbound
protein structure (Eqs.~\eqref{eq:Ebinding} and
\eqref{eq:BreakdownProt}) was not considered in most cases because we
were only interested in differences between complexes. For example,
$\Delta E_{\text{protein deform}}$
(Eq.~\eqref{eq:BreakdownProt}) is simply the energy difference between
the two bound protein structures. In our investigation of amino acid
mutations we approximated $E_{\text{unbound protein}}$ in
Eq.~\eqref{eq:Ebinding} with $E_{\text{bound protein}}$, i.e., the
self-energy of the protein in the bound complex. Again, this
approximation is reasonable as we were only interested in differences
of the binding energy between mutant complexes.

For all computations we used the Amber 9 program with the parm99
force field~\cite{parm99}, and the second implicit water model from
Ref.~\cite{Onufriev04}.  No cut-off was applied.
Hydrogen atoms and the nucleic bases, as well as substituted residues
in our amino acid mutation study, were allowed to rearrange freely to
eliminate steric clashes.  The movement of the protein and the DNA
backbone was restricted by springs with a spring constant of
$1.0~\unit{kcal/(mol\AA^2)}$.
2500 steepest-descent and 2500 conjugate-gradient minimization steps
were applied to each configuration before energies were calculated to
ensure convergence. A typical minimization run for a protein-DNA
complex took about four hours on a 3~Ghz Pentium 4 desktop computer.
%

%Certainly, some relaxation of the structure is necessary and is
%commonly allowed to perform to prevent the energy measurement from
%being dominated by the large energies of overlapping or clashing
%atoms. The stiffer the springs hold the structure, the shorter the
%computational time to reach convergence but the higher the energy
%scales. The looser the springs, the further the structure moves from
%the native state into an energy well, which finite temperatures could
%have prevented. As mentioned in the Introduction others have found
%that the more a structure is allowed to move away from the
%experimental structure, the worse the energy-based methods become at
%predicting binding specificity. Our choice of the spring constant
%keeps the rmsd of the protein backbone around $0.4\AA$. In any case,
%in this work we have uncovered that our method can predict
%\emph{universal} aspects of binding, such as relative binding
%preferences, for which the exact value of the spring constant is
%irrelevant.
While the relaxation of the structures is an essential element of our
method, we cannot allow energy minimization to proceed unhindered.
This is because (1) we do not fully trust the potentials, and (2) the
finite temperature fluctuations (not included) may prevent the
structure moving into certain energy wells. As mentioned in the
Introduction, previous work has indicated that the more the complex is
allowed to move away from the known experimental structure
\cite{Endres04,Morozov05}, the less reliable are the energy-based
methods in predicting binding specificity. The springs introduced in
the previous paragraph limit the drift of the structure, but their
strength is an additional parameter of the problem. In practice, for
the spring {constant} we employ, the rmsd of the protein backbone
changes by about $0.4~\unit{\AA}$ from the native structure.
Fortunately, we find that the relevant aspects of the binding, namely
the relative preferences to different sequences, {are} independent
of the choice of the spring constant{ as long as the structures'
  integrity is preserved. This conclusion was reached after performing
  studies with spring constants of} $1.0, 2.5, 5.0$, and
$7.5~\unit{kcal/(mol\AA^2)}${.}
%We decided that the spring constant
%}$1.0~\unit{kcal/(mol\AA^2)}$ \hl{was the most appropriate one: the
%number of steps until convergence in the energy minimization remained
%few (see Section Materials and Methods), the structure remained close
%to the original x-ray structure, yet the structure could relax
%noticably (rmsd 0.4 A as mentioned).}

%%%%%%%%%%%%%%%%%%%%%%%%%%%%%%%%%%%%%%%%%%%%%%%%%%%%%%%%%%%%%%%%%%%
\section{Results}\label{sec:Results}
\subsection{Comparison with Bioinformatics Scores}\label{sec:Bioinformatics}
In order to assess the quality of binding predictions based on the
all-atom calculations we compared them with predictions made using a
bioinformatic technique
%(PWM method) in their ability to detect cognate sites among decoys.
The PurR transcription factor has been studied extensively and is
therefore particularly well-suited for this task.  Mironov and
co-workers~\cite{Mironov99} compiled a collection of 21 binding sites
to which PurR is considered to bind in {\it E.~coli}.  Assuming
independence of the influence of different base pairs on specificity,
they set up a position-specific weight matrix (PWM) that we use to
calculate bioinformatics scores for various DNA sequences.  Given a
sufficient number of known sites, PWM scores provide a good
approximation of experimentally measured binding energies~\cite{Quake07,Bulyk,Stormo} 
and have sufficient specificity to detect
binding sites in bacterial genomes~\cite{some_Gelfand_Mironov}.

\begin{figure}[h]
\vspace{0.5cm}
\centering
\includegraphics[width=0.75\columnwidth]{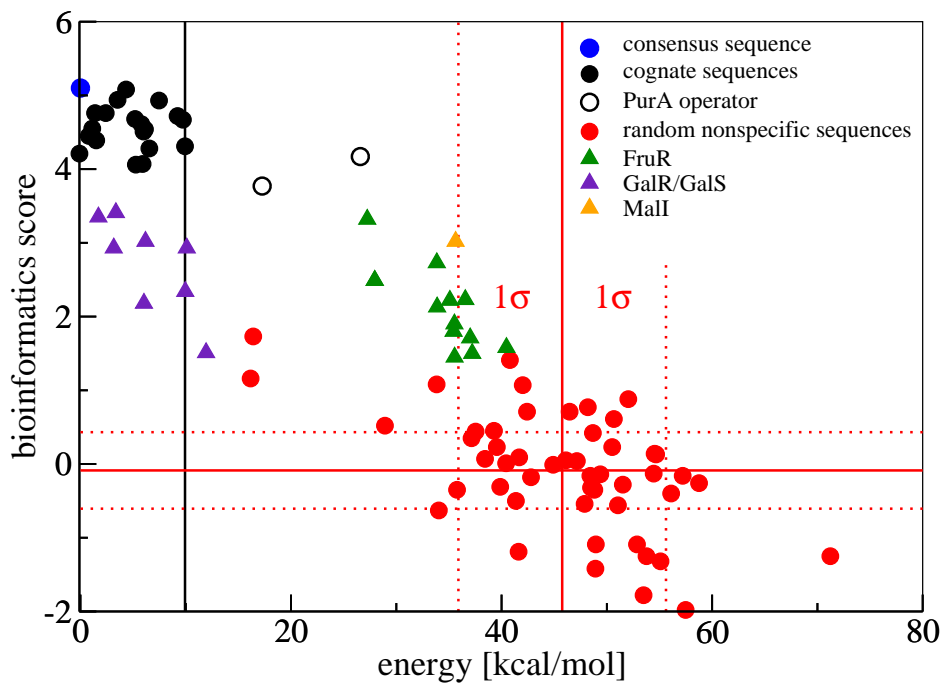}
\caption{Bioinformatics score versus energy. All binding energies are
  shown relative to the binding energy of the consensus sequence
  \seqc~(blue circle) at $0~\unit{kcal/mol}$.  Black circles: 21
  binding sequences selected by Mironov et
  al.~\protect\cite{Mironov99}, PurA sequences are unfilled.  Red
  circles: random non-cognate sequences selected from the {\it
    E.~coli} genome. Green, indigo and orange triangles: FruR,
  GalR/GalS, and MalI operator sites. The solid red lines indicate the
  average energy or average bioinformatics score for the random
  sequences; the dashed lines mark the first standard deviation. The
  solid black line goes through the data point for the third worst
  cognate sequence (a black circle). The two cognate sequences with
  even worse binding energies (hollow black circle) are controversial
  binding sites. {The linear correlation coefficient is }$-0.6${
    for the random sequences and }$-0.8${ for all sequences
    displayed.}  }
\label{fig:Bioinformatics}
\end{figure}

We challenged our structure-based approach, which uses {\em only one known
site} that is a part of the crystal structure, to detect cognate sites
among random ones using the binding energies after minimization.
These energies were also compared to the PWM scores. In
particular, we examined the consensus sequence, the $21$ binding
sequences, $50$ random sequences, and several binding sites of closely
related transcription factors FruR, GalR/GalS, and MalI.  Sites of
homologous transcription factors were chosen because they constitute
particularly challenging sites that are similar to PurR cognate sites
and share the same palindromic structure.
%Figure \ref{fig:Bioinformatics} illustrates the data.
%% \begin{figure}[h]
%% \vspace{0.5cm}
%% \centering
%% \includegraphics[width=7cm]{purr_min_amber.eps}
%% \caption{Bioinformatics score versus energy. All binding energies are
%% shown relative to the binding energy of the consensus sequence
%% \seqc~(blue circle) at $0~\unit{kcal/mol}$.  Black circles: 21 binding
%% sequences selected by Mironov et al.~\protect\cite{Mironov99}, PurA
%% sequences are unfilled.  Red circles: random non-cognate sequences
%% selected from the {\it E.~coli} genome. Green, indigo and orange
%% triangles: FruR, GalR/GalS, and MalI operator sites. The solid red
%% lines indicate the average energy or average bioinformatics score for
%% the random sequences; the dashed lines mark the first standard
%% deviation. The solid black line goes through the data point for the
%% third worst cognate sequence (a black circle). The two cognate
%% sequences with even worse binding energies (hollow black circle) are
%% controversial binding sites.
%% }
%% \label{fig:Bioinformatics}
%% \end{figure}
As shown in Fig.~\ref{fig:Bioinformatics}, bioinformatics scores and
binding energies correlate well {with a linear correlation
  coefficient of }$-0.6${ for the random sequences and }$-0.8${
  for all sequences displayed}. The bioinformatics consensus sequence
has the second lowest energy and random non-cognate sequences are
generally well-separated from cognate sequences. While the separation
between cognate sites and the $50$ random decoys is reassuring, it is
important to find out whether the procedure is able to find cognate
sites among $10^6$ other sites (decoys) on the bacterial genome.

Assuming a Gaussian distribution of the binding energies for random
sites, we can estimate the number of decoys that have binding energies
comparable to the cognate sites.  The distance between the average of
the random sequences (red line) and the third worst cognate site
(black line) is $3.63$~$\sigma$.  (We chose the third worst cognate
site because the two next sequences are the PurA operator sites, see
next paragraph.)  This roughly amounts to one false positive hit in
$7000$ random sites.
Note that 50 random sequences can only yield a rough estimate for this
number.
This number is quite encouraging although it
should only be considered a rough estimate.  For comparison, the
corresponding PWM bioinformatics scores from Ref.~\cite{Mironov99} are
separated by 4.55~$\sigma$ which would amount to 1 false positive hit
in $370,000$ sequences.  (The K12 {\it E.~coli}
genome~\cite{k12_genome} consists of 4.64 Mbps).

The two cognate sequences with the highest energies are the two PurA
operators (unfilled black circles).  Indeed, the suggestion that the
PurA operon may be regulated by PurR is controversial (see
Refs.~\cite{Mironov99,Meng90,He94}).  Although at the lower end of the
spectrum, the bioinformatic scores for these sites are comparable with
other cognate sites, while our computations give distinctly higher
binding energies.
%
%\cite{He94} and the feature
%of having two binding sites we suggest may be related to them having
%relatively high binding energies. In fact, Mironov and co-workers
%point out that the binding of PurR to the operators is
%``questionable''\cite{Mironov99}, a conclusion that they base on the
%work of Meng et al.\cite{Meng90} and He et al.\cite{He94}. Our
%all-atom method is able to detect this anomaly.
%

Testing the energy-based approach on operator sites regulated
by other members of the LacI family is a particularly challenging task.
Although the FruR and MalI binding sequences are energetically well
separated from the PurR cognate sequences, GalR/GalS binding
sequences are not. (The bioinformatics score appears to have 
less difficulty with these sites.)
%Yet, bioinformatics scoring is able to distinguish these two sets. 
%Although one may interpret this as a failure of
%all-atom sequence predictions, we wonder whether there may be a
%connection between PurR and GalR/GalS thus far overlooked.

Finally, we would like to point out that the absolute energy scale is incorrect,
in line with the conclusions of Ref.~\cite{Paillard04-1}.
Excluding PurA, the range of binding energies for cognate 
sequences is $10~\unit{kcal/mol}$ which is clearly too large.
%This was already indicated in
%Ref.~\cite{Paillard} which compared energy differences of DNA point
%mutations with differences in the experimental free energies. 
The underlying assumptions and approximations of the method are, 
however, quite considerable and quantitative agreement cannot really be expected.

\subsection{Direct and Indirect Contributions to the Binding Energy, Sequence Logos}\label{sec:Breakdown}

The binding free energies can be subdivided into two parts: 
Direct interactions between the transcription factor and
DNA, and indirect contributions due to sequence-specific DNA bending. 
In recent work, Paillard and Lavery~\cite{Paillard04-1} noted that the level of
each contribution varies significantly from complex to complex.
Their method is based on a careful analysis of a subset of sequences
with particularly low binding energies, after having computed the 
energies of all possible $4^{N}$ sequences. Here, we 
propose a simple method which can distinguish between contributions
of bending and protein-DNA interactions on the basis of a rather limited
set of measurements.

%Simple bioinformatics sequence specificity prediction techniques are
%based on the assumption of independent contributions of the base pairs
%to the sequence specificity. At each position of the DNA sequence a
%weight is given to each of the four possible base pairs. The
%bioinformatics score for binding of a given sequence then is the sum
%of the weights of the sequence's base pairs at each position. We
%decided to analyze whether this approach can be supported and also
%used by our all-atom methods.

To understand the source of the sequence
specificity of PurR, we partitioned its binding energy as follows:
\begin{equation}
\begin{split}
E_{\text{binding}} & = E_{\text{interaction}} \\
                   & + E_{\text{DNA deform}} \\
                   & + E_{\text{protein deform}}
\end{split},
\end{equation}\label{eq:Breakdown}
where
\begin{equation}
\begin{split}
E_{\text{interaction}} & = E_{\text{protein-DNA complex}} \\
                       & - E_{\text{bound protein}} \\
                       & - E_{\text{bound DNA}}
\end{split},\label{eq:BreakdownInt}
\end{equation}
\begin{equation}
\begin{split}
E_{\text{DNA deform}} & = E_{\text{bound DNA}} \\
                   & - E_{\text{straight DNA}}
\end{split},\label{eq:BreakdownDNA}
\end{equation}
and
\begin{equation}
\begin{split}
E_{\text{protein deform}} & = E_{\text{bound protein}} \\
                           & - E_{\text{unbound protein}}
\end{split}.\label{eq:BreakdownProt}
\end{equation}

We next computed direct and indirect contributions for both
cognate and random sequences and compare the differences: On average,
$E_{\text{interaction}}$ was lower by $34~\unit{kcal/mol}$ for the
cognate sites compared to random ones, and $E_{\text{DNA deform}}$ lower
by $7~\unit{kcal/mol}$.  Assuming that the force field reproduces the
correct {\em ratios} of direct and indirect contributions, specificity
towards PurR is predominantly determined by protein-DNA
interactions. It is interesting to note that $E_{\text{protein
deform}}$ was slightly higher for cognate sites ($2~\unit{kcal/mol}$)
indicating that the interactions were strong enough to bend the
protein towards a slightly unfavorable position.

To study the contribution of individual base pairs to specificity, and
to significantly speed up computations, we used energy minimization to
calculate a position-specific energy matrix, analogous to
PWM~\cite{Mironov99}.  As specificity towards PurR is dominated by
direct, pairwise interactions, we computed the change in
$E_{\text{interaction}}$ due to each possible single mutation of the
consensus sequence and set up an energy-based weight matrix EBWM
(Table \ref{tbl:WeightMatrix}a).
(While it is in principle possible to construct a statistical weight
matrix based on the top 21 sites identified by energy minimization,
this would effectively reproduce the experimental PWM of
Ref. \cite{Mironov99}.)
The interaction energy for an
arbitrary DNA sequence can now be computed by adding the appropriate
base pair energies. This requires only a limited number of
computations at the cost of being less accurate.
%
%While it is in principle possible to construct a statistical weight
%matrix based on the top 21 sites identified by energy minimization,
%this would effectively reproduce the experimental PWM of
%Ref. \cite{Mironov99}. We chose instead to construct an energy-based
%weight matrix starting from the consensus sequence, and making point
%mutations at each site.

Although computationally efficient, both EBWM and PWM methods are based
on the assumption that the contributions of individual base pairs are
independent from each other, neglecting many-body effects, such as due to
solvation. EBWM calculated using $E_{\text{interaction}}$ also ignores
sequence-dependent contributions of DNA deformation to the binding energy. 
Can we improve upon this by taking the bending energy into account?

The DNA bending energy cannot be easily decomposed into contributions
of individual base pairs; the energetic contribution of a single base
pair to DNA bending not only depends on its neighbors but on the whole
sequence.  This makes the exact treatment of the problem
computationally challenging.  Lavery and coworkers~\cite{Lafontaine01,Paillard04-1,Paillard04-2} address this issue by constructing an
'average structure' into which they can substitute all possible
sequences and create a sequence logo based on the sequences with the
lowest energy. Endres et al.~\cite{Endres04,Endres06} employ an
efficient scheme to screen sequences and only compute those which look
promising. Thus they afford to compute bending energies for each
individual sequence.

%of independent contributions of the base pairs
%to sequence specificity. At each position of the DNA sequence a
%weight is given to each of the four possible base pairs. The
%bioinformatics score for binding of a given sequence then is the sum
%of the weights of the sequence's base pairs at each position. 
%We decided to analyze whether this approach can be supported and also
%used by our all-atom methods.
%
%The protein-DNA interaction
%energy has the property that, in principle, it is simply the sum of
%the interaction energy of each base pair with the protein. However, in
%practice, since our molecules are solvated, many-body effects play a
%role and retain some influence between neighboring base pairs, which
%we shall ignore. So, we computed the change in
%$E_{\text{interaction}}$ due to each possible single mutation of the
%consensus sequence and set up a weight matrix, which is a look-up
%table from which the interaction energy $E_{\text{interaction}}$ for
%any DNA sequence with the protein can be computed by adding the
%appropriate base pair interaction energies. By repeating the energy
%minimization for each DNA mutant separately, we introduced another
%non-additive energy component because each DNA mutant relaxes to a
%slightly different structure. But the error analysis shows that the
%influence of these many-body effects is negligible.

We sought to improve the EBWM approximation by making a ``zeroth
order'' estimate of the bending energy.  The bending energy can also
be subdivided into two parts: interactions between base pairs and the
backbone, which are approximately independent of the other base pairs,
and interactions between nucleic bases, which are not. Nevertheless,
we tested if the results can be improved by including an additive
bending term to $E_{\text{interaction}}$. Such a treatment is
tantamount to considering interactions of single base pairs in the
``mean-field'' environment of the consensus sequence.
Similar to the
case of $E_{\text{interaction}}$, we computed the change in
$E_{\text{DNA deform}}$ due to each possible singe base pair
substitution in the consensus sequence and set up a second EBWM
(Table~\ref{tbl:WeightMatrix}b).
Thus, for every position along the DNA sequence, the change in bending
energy due to a point mutation is measured. Then, the total bending
energy for an arbitrary sequence is approximated as the sum of the
changes in bending energy at each position. This approximation only
captures the interactions of the base pairs and the backbone and some
mean-field portion of the interaction between base pairs, but leaves
out base stacking energies which are explicitly not pairwise additive.
More precisely, to find $E_{\text{DNA deform}}$, as defined in
Eq. (\ref{eq:BreakdownDNA}), $E_{\text{bound DNA}}$ and
$E_{\text{straight DNA}}$ are needed. The former is computed by
excising the DNA from the energy minimized protein-DNA complex
structure and measuring its energy without the surrounding
protein. $E_{\text{straight DNA}}$ is, of course, the energy of the
DNA in its free form. Subtracting the two energies yields the energy
of deformation, $E_\text{DNA deform}$.

%Clearly, a
%bending energy cannot really be assigned to each base pair.
% at each
%position in the DNA independently from the others. 
%The bending energy
%of some base pair at some position depends not only on which base pair
%it is but also on the base pairs on all other positions within some
%reasonable distance. However, how would one fair if one assumed some
%mean-field interaction with the other base pairs and, in fact,
%approximated the bending energy as a sum of contributions from each
%position of the DNA? We computed, similar to the case of
%$E_{\text{interaction}}$, the change in $E_{\text{DNA-DNA}}$ due to
%each possible single mutation of the consensus sequence and set up a
%weight matrix.

\begin{table}[h]
\subtable[$\Delta E_{\text{interaction}}$]{
\begin{tabular} {|l | l | l| l|}
\hline
A & C & G & T \\
\hline
0.0 & -0.5 & -0.7 & -0.7 \\
1.3 & 0.0 & 0.4 & -0.6 \\
4.7 & 14.3 & 0.0 & 8.4 \\
-1.0 & 0.0 & 2.3 & 1.2 \\
0.0 & 2.1 & 3.7 & 3.4 \\
0.0 & 3.5 & 3.4 & 3.8 \\
0.0 & 2.0 & 0.6 & 2.1 \\
6.2 & 0.0 & 0.1 & 5.2 \\
5.3 & 0.1 & 0.0 & 6.0 \\
2.2 & 0.6 & 2.1 & 0.0 \\
3.9 & 3.4 & 3.6 & 0.0 \\
3.4 & 3.8 & 2.1 & 0.0 \\
0.9 & 2.3 & 0.0 & -0.9 \\
8.4 & 0.0 & 14.4 & 4.6 \\
-0.6 & 0.6 & 0.0 & 1.4 \\
-0.7 & -0.6 & -0.6 & 0.0 \\
\hline
\end{tabular}
\label{tbl:WeightMatrixInt}
}
\subtable[$\Delta E_{\text{interaction}} + \Delta E_{\text{DNA deform}}$]{
\begin{tabular} {|l | l | l| l|}
\hline
A & C & G & T \\
\hline
0.0 & 0.3 & -0.6 & 0.2 \\
1.9 & 0.0 & 0.0 & 0.4 \\
7.6 & 19.6 & 0.0 & 15.1 \\
2.3 & 0.0 & 2.4 & 0.1 \\
0.0 & 1.6 & 2.2 & 4.2 \\
0.0 & 4.4 & 4.5 & 5.7 \\
0.0 & 0.5 & 2.8 & 0.4 \\
2.3 & 0.0 & -0.6 & 3.3 \\
3.8 & -0.8 & 0.0 & 1.4 \\
0.4 & 2.7 & 0.3 & 0.0 \\
8.4 & 4.8 & 6.7 & 0.0 \\
3.3 & 2.4 & 0.3 & 0.0 \\
-0.2 & 2.1 & 0.0 & 2.2 \\
14.9 & 0.0 & 19.3 & 7.8 \\
-0.1 & 0.2 & 0.0 & 1.8 \\
-0.2 & -0.5 & -0.2 & 0.0 \\
\hline
\end{tabular}
\label{tbl:WeightMatrixBoth}
}
\caption{Position specific energy matrices based on: a) direct
interaction energies, and b) interaction energies plus bending
corrections. The energies are normalized to the consensus sequence,
which has, accordingly zero binding energy. This is why a ``$\Delta$''
appears in front of the energies. All contributions from each base
pair (including bending) were considered to be independent of the
other base pairs. Energies are given in units of $[\unit{kcal/mol}]$.
Only the first decimal place is shown.}
\label{tbl:WeightMatrix}
\end{table}

In an effort to gauge the usefulness of the two matrices, we computed
the energy difference between the worst cognate sequence and all the
sequences encountered in scanning the {\it E. coli} genome excluding
the PurA operator sites. We repeated this to find the energy
difference between the worst cognate sequence and the 50 nonspecific
random sequences discussed above. If only protein-DNA interactions are
taken into account, the separation between the lowest and the average
of all sequences is 2.74~$\sigma$ (3.2~$\sigma$ based on the 50
nonspecific sequences). If the additive bending correction is included
the distance is also 2.74~$\sigma$ (3.38~$\sigma$ based on the 50
nonspecific sequences). The 50 nonspecific random sequences are too
few to allow reliable conclusions; clearly, the genome scan is more
significant.
%Assuming a Gaussian distribution,
%this number corresponds to 1 false positive hit in approximately 300
%sequences.
%
These results indicate that a simplified EBWM approach which only
considers additive interactions is not sufficient to provide accurate
discrimination of sites.  Similarly, we expect the that experimentally
obtained PWM based on $\Delta G_\text{binding}$ of single base pair
mutants of the consensus sequences~\cite{Mironov99} to suffer from a
similar lack of discrimination power.
%If energy minimizations are applied to determine weight matrices or consensus 
%sequences, more efficient schemes, which are able to calculate all \cite{Lafontaine01, Paillard04-2, Paillard04-1} 
%or at least the most relevant sequences \cite{Endres04, Endres06}, need to be implemented.

%The errors introduced by these additivity approximations were
%calculated by computing $E_{\text{interaction}}$ and
%$E_{\text{DNA-DNA}}$ by way of the two weight matrices and comparing
%them to the direct measurements of the energies using the
%structures. Since the weight matrices are based on mutations of the
%consensus sequence one expects the energies of the specific
%sequences, which are similar to the consensus sequence, to contain
%small errors and the random sequences to contain larger errors.
%[INSERT DATA ABOUT ERRORS.]
%
%Scanning the entire E. Coli genome [reference?] with the weight
%matrices one obtains the $E_{\text{interaction}}$ and
%$E_{\text{interaction}}+E_{\text{DNA-DNA}}$ distributions of all
%sequences.
%[INSERT COMMENTS.]

\begin{figure}[ht]
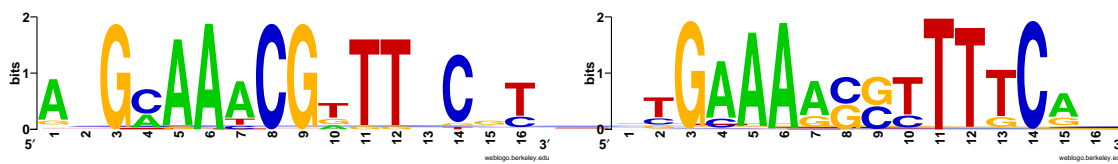
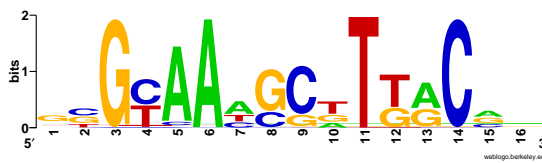

  \centering
  \subfigure[~Bioinformatics logo from Ref.~\cite{Mironov99},
  based on the sequences of 21 experimentally known binding sites.]
  {\label{fig:edge-a}\includegraphics[width=7.5cm]{21bindingseqlogo.eps}}
  \subfigure[~$E_{\text{interaction}}$-based Logo, obtained from the Boltzmann probabilities of residues from
  site specific interaction energies listed in Table~\ref{tbl:WeightMatrix}a).]
  {\label{fig:edge-b}\includegraphics[width=7.5cm]{matrixinteractionEturnedintoseq.eps}}
  \subfigure[~($E_{\text{interaction}}+E_{\text{DNA deform}}$)-based
  Logo,  obtained from the Boltzmann probabilities of residues from
  site specific interaction energies listed in  Table~\ref{tbl:WeightMatrix}b). 
  This includes an estimate of the bending energy of the DNA as described in the text.]
  {\label{fig:edge-c}\includegraphics[width=7.5cm]{matrixsuminteractionbendingturnedintoseq.eps}}
  \caption{Consensus Sequence Logos.}
  \label{fig:Logos}
\end{figure}

To visualize the contribution of individual base pairs to the PurR
motif, and hence to the specificity of recognition, we converted the
information contained in the EBWM into a sequence logo (Figure
\ref{fig:Logos}).  This is done by using Boltzmann weights to
represent the frequency of occurrence of each base pair at each
position. Room temperature was used in the Boltzmann factors, that is,
$kT=0.59~\unit{kcal/mol}$.  Comparison with the bioinformatics logo
from Ref.~\cite{Mironov99} (Fig.~\ref{fig:Logos}a) indicates that the
structure-based method is able to reproduce the specificity of most of
the positions in the PurR motif.  In particular, base pairs at
positions 3 (G), 5 (A), 6 (A), 11 (T), 12 (T), and 14 (C) are
identified correctly, but the method cannot distinguish between the
consensus CpG versus GpC in positions 8 and 9.  Base pairs 8 and 9
play an important role in the binding of DNA because PurR bends DNA by
intercalating a lysine side chain between these two base
pairs.~\cite{Schumacher94} Either CpG is selected in nature for
reasons that cannot be explained by binding energy considerations or
the force field cannot capture a subtle difference in binding between
CpG and GpC in the center of the binding sequence. Nevertheless, it is
surprising that although the computed energy differences are too big
compared to the experimental energy differences, the sequence logo is
recreated rather accurately.

The Boltzmann weights of each base pair at any position along the DNA
represent a probability distribution for the four base pairs at that
position. Two different probability distributions $p(i)$ and $q(i)$
can be compared using the relative entropy measure (Kullback-Leibler
divergence \cite{Kullback51}), $\sum_i p(i) \log p(i)/q(i)$. Here,
$q(i)$ is the probability distribution of the base pairs $i=\text{AT,
CG, GC, or TA}$ at some position in the DNA derived from the
bioinformatics weblogo and $p(i)$ is the probability distribution for
the base pairs at the same position derived from the Boltzmann weight,
which is computed with our method. The more dissimilar two probability
distributions are, the larger is their relative entropy. Excluding the
two end base pairs and the two middle base pairs with the problematic
CpG ambiguity, the distance between the probability distribution based
on $E_\text{interaction}$ and the bioinformatics probability
distribution is, on average, $0.8$. Excluding the same base pairs, the
distance between the probability distribution based on
$E_\text{interaction}+E_\text{DNA deform}$ and the one based on
bioinformatics is $0.3$, which reflects a clear {improvement}.

%% \begin{figure}[ht]
%%   \centering
%%   \subfigure[~Bioinformatics logo from Ref.~\cite{Mironov99},
%%   based on the sequences of 21 experimentally known binding sites.]
%%   {\label{fig:edge-a}\includegraphics[width=7.5cm]{21bindingseqlogo.eps}}
%%   \subfigure[~$E_{\text{interaction}}$-based Logo, obtained from the Boltzmann probabilities of residues from
%%   site specific interaction energies listed in Table~\ref{tbl:WeightMatrix}a).]
%%   {\label{fig:edge-b}\includegraphics[width=7.5cm]{matrixinteractionEturnedintoseq.eps}}
%%   \subfigure[~($E_{\text{interaction}}+E_{\text{DNA deform}}$)-based
%%   Logo,  obtained from the Boltzmann probabilities of residues from
%%   site specific interaction energies listed in  Table~\ref{tbl:WeightMatrix}b). 
%%   This includes an estimate of the bending energy of the DNA as described in the text.]
%%   {\label{fig:edge-c}\includegraphics[width=7.5cm]{matrixsuminteractionbendingturnedintoseq.eps}}
%%   \caption{Consensus Sequence Logos.}
%%   \label{fig:Logos}
%% \end{figure}

In summary, we have shown that: (i) the contribution of indirect
readout due to DNA bending is significant; (ii) this contribution
cannot be easily accounted for by a site-specific approximation;
(iii) the EBWM provides a fast way of estimating the binding energy
but suffers from a significant loss of statistical power; (iv) a
structure-based energy calculation is able to capture most of the PurR
motif, but fails to identify the central base pairs correctly. 
This suggests a hybrid strategy of first using EBWMs to scan for potential 
binding sites, and then following up by a more computationally intensive energy 
minimization for these candidates.

\subsection{Investigation of DNA and Protein Mutants}
\label{sec:Mutations}

In this section we  compare  the binding energies derived from
the structure-based approach with the experimental free energies of
binding for a number of DNA and amino acid point
mutations.~\cite{Glasfeld99,Arvidson98} 
In particular, we investigated the following
sequences bound to the protein wild type and K55A mutant:
%Thus far we have shown that atomistic energy calculations can measure
%up to the more widespread bioinformatics techniques. However, they can
%also reach beyond statistical analyses. 
%We studied the change in
%$E_{\text{binding}}$ due to mutations of the PurR protein, which we
%compared with experimental studies. We introduced single point
%mutations to yield the PurR mutants a)~K55A, b)~L54M, c)~L54S,
%d)~L54T, and e)~L54V. XnY, by convention, stands for the mutant in
%which the amino acid with one letter abbreviation X at position n is
%replaced by the amino acid Y.
\begin{equation}
\begin{tabular}{ l l }
\seqc = & acgcaa(a)cg(t)ttgcgt (consensus),\\
\seqone = & acgcaa(c)cg(g)ttgcgt,\\
\seqtwo = & acgcaa(g)cg(c)ttgcgt,\\
\seqthree = & acgcaa(t)cg(a)ttgcgt.\\
\end{tabular}\nonumber
\end{equation}
In addition, we studied protein mutants L54M, L54S, L54T, and L54V bound to the
consensus sequence. The results are summarized in Table \ref{tbl:Mutations}.

\begin{table}[h]
\begin{tabular} {|l | l | l|}
\hline
\multicolumn{3}{|l|}{\centerline{Binding to wild-type PurR}} \\
\hline
DNA Sequence & $\Delta E_{\text{binding}}$ & $\Delta G_{\text{binding}}$ (experiment) \\
\seqc & 0 & 0.0 \\
\seqthree & 0.16 & 0.8 \\
\seqone & 2.02 & 1.6 \\
\seqtwo & 6.78 & 3.2 \\
\hline
\multicolumn{3}{|l|}{\centerline{Binding to the K55A mutant}} \\
\hline
DNA Sequence & $\Delta E_{\text{binding}}$ & $\Delta G_{\text{binding}}$ (experiment) \\
\seqone & -7.53 & -0.06 \\
\seqtwo & -4.47 & -0.46 \\
\seqc & 0 & 0 \\
\seqthree & 1.13 & 0.5 \\
\hline
\multicolumn{3}{|l|}{\centerline{PurR mutants bound to the consensus sequence}} \\
\hline
Mutant & $\Delta E_{\text{binding}}$ & $\Delta G_{\text{binding}}$ (experiment) \\
WT & 0 & 0 \\
L54M & 5.79 & 0.38 \\
L54S & 16 & larger, not measured \\
L54T & 10.05 & ,, \\
L54V & 6.15 &  ,, \\
K55A & 12.55 & 3.48 \\
\hline
\end{tabular}
\caption{Calculated changes in binding energies of DNA and amino acid point
mutations compared with experiments~\cite{Glasfeld99,Arvidson98}. 
When only the DNA is mutated, the
binding order is correct (top panel). When both DNA and the
protein are mutated (middle panel) two DNA mutants are lower in
binding energy and one higher than the original
sequence. This is correctly identified by our method, but the
binding preference to \seqone~and \seqtwo~is reversed. When only the
protein is mutated, the binding preferences of the DNA to the
mutants are correctly captured (bottom panel). Energies are
given in kcal/mol and measured relative to the respective consensus
protein-DNA complex.}
\label{tbl:Mutations}
\end{table}

Although the overall energy scale is incorrect, we are able to reproduce (with one
exception) the correct order of experimental binding free energies for
all DNA and amino acid mutants. Qualitatively similar results were
obtained in Ref.~\cite{Paillard04-1}, which, however, only considered
mutations of the DNA sequence.  
The sampling of amino acid mutants is
particularly relevant because it allows us to predict whether simple modifications
of transcription factors can lead to higher or lower binding affinity.

%The wild-type protein's experimental results correlate well with our
%calculations. The scale though is off by a factor of about 2. For the
%K55A mutant protein our method detects correctly that \seqone and
%\seqtwo are preferred to the consensus sequence \seqc and that \seqc
%is preferred to \seqthree. However, the order of preference of \seqone
%and \seqtwo is erroneously switched. The L54M mutant binds the
%consensus sequence less strongly than the wild-type protein does. But
%the magnitude of the effect of the mutation is grossly
%exaggerated. The authors of the experimental study do not provide data
%for the other three mutations but mention that binding was much less
%tight, which our data reflects appropriately as higher energies of
%binding for those three mutants. Finally, the relative binding
%energies of the three mutants to the consensus sequence \seqc also
%correlate with the experimental relative binding free energies, the
%trend is correct, yet, again, the scale is off.

%%%%%%%%%%%%%%%%%%%%%%%%%%%%%%%%%%%%%%%%%%%%%%%%%%%%%%%%%%%%%%%%%%%
\section{Discussion}\label{sec:Discussion}

It is clearly desirable to understand protein-DNA binding on a molecular level,
and all-atom energy calculations based on minimizing experimental 
structures are a promising step towards this goal. 
In this work we studied the feasibility of predicting the affinity of
a transcription factor to different sequences, by using off-the-shelf
and widely used interaction potentials.
Our main goal was to test whether computation of energies using
such a potential allows discrimination of cognate sites from random decoys.
Using the example of the PurR transcription factor as a model
system, and starting from the structure this protein bound to a specific 
DNA sequence, we tested our  method in the following ways:
\begin{itemize}
\item examined its ability to deliver lower binding energies to cognate sites as
  compared to random decoys;
\item estimated the number of random sites that have binding energies
   comparable to cognate sites, thereby assessing 
  the potential of this method to detect sites in long genomic
  sequences;
\item compared the performance of this structure-based method with the
  bioinformatic PWM technique that requires {\it a priori} knowledge
  of {\em several} cognate sites;
\item examined PurR motifs obtained using structure-based calculations,
  and compared them with the motif inferred from the cognate sites;
\item calculated the change in the binding energy due to mutations in
  the protein and DNA and compared with experimentally measured
  $\Delta G_\text{binding}$.
\end{itemize}
We further investigated the contribution of the sequence-dependent DNA
bending and tested whether computations can be accelerated using
an EBWM approach. This systematic and diverse testing makes our study
complimentary to other recent works~\cite{Donald07,Lafontaine01,Paillard04-1,Paillard04-2,Endres04,Morozov05,Liu07}.

Overall, the changes in energy of the minimized structures correlate well with
corresponding bioinformatics scores and are accurate enough to
discriminate between binding and random sequences. 
Unfortunately, they are not sufficiently discriminating to enable
systematic scanning of entire genomes. 
The method can, however, distinguish between weak and strong
binding sites and, to a lesser extent, between operator sites of
related factors.  
%In this context, we propose to use the measured energies to differentiate between contributions of direct protein-DNA interactions and indirect contributions due to bending.

To highlight the contributions of individual base pairs, we compared a
motif logo obtained using structure-based calculations with the logo
for cognate sites. While most of the positions reflect the cognate
motif correctly, the two central base pairs are predicted incorrectly with
atomistic force fields- indicating no difference between G and C in
these positions.  
This difficulty is likely due to a complicated binding
mechanism through lysine intercalation used by PurR to bind the
central base pairs. It is possible that sequence-dependent bendability
of DNA makes CpG a preferred base pair in the center of a sharply bent
PurR site.  
Understanding the molecular mechanism of recognition of
the central base pairs requires further studies using molecular
dynamics.  A method that can resolve this discrepancy is likely to
provide a significant improvement to structure-based predictions for
PurR and other transcription factors that bend DNA.

Our analysis provides a glimpse into the promise of structure- and
interaction-based methods.  Relatively crude computations are able to
predict the correct {\em order} of binding energies. This is
particularly useful for the study of amino acid mutants which cannot
be investigated with standard bioinformatics methods.  There are
several reasons for the limited success of our approach.  First, the
force fields employed are likely not accurate enough to deliver
precision of the binding energy at the level of a few $\unit{kcal/
mol}$, as required to discriminate cognate sites, especially when an
implicit water model is used. {(The ParmBSC0 force field, for
example, could be used in the future because of its improved
treatment of non-canonical backbone conformations compared to
PARM99. }\cite{Perez07}{)} Second, our procedure crudely
approximates the differences in the binding free energy by ignoring
entropy contributions and by limiting the flexibility of the protein-
DNA complex through the use of fast energy minimization, thus not
allowing for major rearrangements of the structure.
%
%So, the restraints on the protein and the DNA, which keep them from
%deviating too far from the native experimental structure, could be
%increasing the energy scales in the system, thus exaggerating the
%energy differences.
{
Furthermore, the restraints on the protein and the DNA, which keep
them from deviating too far from the native experimental structure,
could be increasing the energy scales in the system, thus exaggerating
the energy differences. The energies of the restraining springs are
not included in the calculations, but the springs may keep the
structure from relaxing to the equilibrium coordinates that the force
field favor.}
The (qualified)
success of this simple approach suggests that further optimization of
the force field \cite{Donald07} and conformational sampling
(e.g. similar to those of \cite{Endres04}) may lead to significant
improvements.  Resolution of these issues is necessary to gain a
better quantitative understanding of protein-DNA binding.

Structure-based methods are more laborious than bioinformatics but
less costly and elaborate than experiments. Although accuracy is still
somewhat lacking in current implementations,  the results are promising and
still leave considerable room for improvements. 
%In future studies effects of temperature, fluctuations, and entropy need to be addressed.
%Nevertheless, present results are already quite encouraging. 
%Structure 
%based methods are accurate enought to discrimiate between binding and random
%sequences, but not to a point that the genome can be scanned systematically
%for new sites. 
Promising applications which already appear feasible include the study
of sequence dependent motion of proteins along DNA, and investigations
of simple amino acid point mutations in conjunction with experiments.

\section{Acknowledgment} 

This work was supported by the National Science Foundation Grant
DMR-08-03315 and by the Deutsche Forschungsgemeinschaft grant
VI237/1. LM acknowledges generous support by NEC fund and by the
National Center for Biomedical Computing {\it i2b2}. PV would like to
acknowledge the John von Neumann Institute for Computing in J\"ulich
for providing computational resources.

%\bibliographystyle{phreport} %\bibliography{biblio_febr}
%\bibliographystyle{nar}
%\bibliography{article}

%\begin{thebibliography}{10}
%\end{thebibliography} 

\clearpage

\end{document}